\documentclass[aps,prl,twocolumn,superscriptaddress]{revtex4}
\usepackage{amssymb}
\usepackage{amsfonts}
\usepackage{graphicx}
\usepackage{amsmath}
\usepackage{array}
\begin{document}


\title{Single-crystal growth and extremely high $H_{c2}$ of 12442-type Fe-based superconductor KCa$_2$Fe$_4$As$_4$F$_2$}

\author{Teng Wang} \affiliation{State Key
Laboratory of Functional Materials for Informatics, Shanghai
Institute of Microsystem and Information Technology, Chinese Academy
of Sciences, Shanghai 200050, China}\affiliation{CAS Center for Excellence in Superconducting
Electronics(CENSE), Shanghai 200050, China}\affiliation{School of Physical Science and Technology, ShanghaiTech University, Shanghai 201210, China}

\author{Jianan Chu}
\affiliation{State Key Laboratory of Functional Materials for
Informatics, Shanghai Institute of Microsystem and Information
Technology, Chinese Academy of Sciences, Shanghai 200050,
China}\affiliation{CAS Center for Excellence in Superconducting
Electronics(CENSE), Shanghai 200050, China}\affiliation{University
of Chinese Academy of Sciences, Beijing 100049, China}

\author{Hua Jin}
\affiliation{State Key Laboratory of Functional Materials for
Informatics, Shanghai Institute of Microsystem and Information
Technology, Chinese Academy of Sciences, Shanghai 200050, China}\affiliation{CAS Center for Excellence in Superconducting
Electronics(CENSE), Shanghai 200050, China}

\author{Jiaxin Feng}
\affiliation{State Key Laboratory of Functional Materials for
Informatics, Shanghai Institute of Microsystem and Information
Technology, Chinese Academy of Sciences, Shanghai 200050, China}\affiliation{CAS Center for Excellence in Superconducting
Electronics(CENSE), Shanghai 200050, China}
\affiliation{University of Chinese Academy of Sciences, Beijing 100049, China}

\author{Lingling Wang}
\affiliation{State Key Laboratory of Functional Materials for
Informatics, Shanghai Institute of Microsystem and Information
Technology, Chinese Academy of Sciences, Shanghai 200050,
China}

\author{Yekai Song}
\affiliation{State Key Laboratory of Functional Materials for
Informatics, Shanghai Institute of Microsystem and Information
Technology, Chinese Academy of Sciences, Shanghai 200050, China}\affiliation{CAS Center for Excellence in Superconducting
Electronics(CENSE), Shanghai 200050, China}
\affiliation{University of Chinese Academy of Sciences, Beijing 100049, China}

\author{Chi Zhang}
\affiliation{State Key Laboratory of Functional Materials for
Informatics, Shanghai Institute of Microsystem and Information
Technology, Chinese Academy of Sciences, Shanghai 200050, China}\affiliation{CAS Center for Excellence in Superconducting
Electronics(CENSE), Shanghai 200050, China}
\affiliation{University of Chinese Academy of Sciences, Beijing 100049, China}

\author{Wei Li}
\affiliation{State Key Laboratory of Surface Physics and Department of Physics, Fudan University, Shanghai 200433, China}
\affiliation{Collaborative Innovation Center of Advanced Microstructures, Nanjing 210093, China}

\author{Zhuojun Li}
\affiliation{State Key Laboratory of Functional Materials for
Informatics, Shanghai Institute of Microsystem and Information
Technology, Chinese Academy of Sciences, Shanghai 200050, China}\affiliation{CAS Center for Excellence in Superconducting
Electronics(CENSE), Shanghai 200050, China}

\author{Tao Hu}
\affiliation{State Key Laboratory of Functional Materials for
Informatics, Shanghai Institute of Microsystem and Information
Technology, Chinese Academy of Sciences, Shanghai 200050, China}\affiliation{CAS Center for Excellence in Superconducting
Electronics(CENSE), Shanghai 200050, China}

\author{Da Jiang}
\affiliation{State Key Laboratory of Functional Materials for
Informatics, Shanghai Institute of Microsystem and Information
Technology, Chinese Academy of Sciences, Shanghai 200050, China}\affiliation{CAS Center for Excellence in Superconducting
Electronics(CENSE), Shanghai 200050, China}

\author{Wei Peng}
\affiliation{State Key Laboratory of Functional Materials for
Informatics, Shanghai Institute of Microsystem and Information
Technology, Chinese Academy of Sciences, Shanghai 200050, China}\affiliation{CAS Center for Excellence in Superconducting
Electronics(CENSE), Shanghai 200050, China}

\author{Xiaosong Liu}
\affiliation{State Key Laboratory of Functional Materials for
Informatics, Shanghai Institute of Microsystem and Information
Technology, Chinese Academy of Sciences, Shanghai 200050, China}\affiliation{CAS Center for Excellence in Superconducting
Electronics(CENSE), Shanghai 200050, China}\affiliation{School of Physical Science and Technology, ShanghaiTech University, Shanghai 201210, China}

\author{Gang Mu}
\email[]{mugang@mail.sim.ac.cn} \affiliation{State Key Laboratory of
Functional Materials for Informatics, Shanghai Institute of
Microsystem and Information Technology, Chinese Academy of Sciences,
Shanghai 200050, China}\affiliation{CAS Center for Excellence in Superconducting
Electronics(CENSE), Shanghai 200050, China}

\begin{abstract}
Millimeter sized single crystals of KCa$_2$Fe$_4$As$_4$F$_2$ were grown
using a self-flux method. The chemical compositions and crystal structure were
characterized carefully. Superconductivity with the critical transition $T_c$ = 33.5 K
was confirmed by both the resistivity and magnetic susceptibility
measurements. Moreover, the upper critical field $H_{c2}$ was studied by the resistivity measurements under 
different magnetic fields. A rather steep increase for the
in-plane $H_{c2}^{ab}$ with cooling, $d\mu_0H_{c2}^{ab}/dT$$\mid$$_{T_c}$ = -50.9 T/K, was observed, indicating an extremely high upper critical field. 
Possible origins for this behavior were discussed.
The findings in our work is a great promotion both for understanding the physical properties and applications of 12442-type Fe-based superconductors.

\end{abstract}

\pacs{74.20.Rp, 74.25.Ha, 74.70.Dd} \maketitle

\section{I. Introduction}
The Fe-based superconductors (FeSCs)~\cite{LaFeAsO} share at least one common feature with the cuprate superconductors~\cite{Bednorz1986}: the crystal structure consists of insulating layers serving as the 
carrier reservoir and conducting layers which are the key section for the superconductivity. Meanwhile, the subtle difference is also noticed by the material scientists. 
For the cuprates, superconductivity can be observed in systems with monolayers~\cite{Bednorz1986,Michel1987}, bilayers~\cite{Maeda1988},
trilayers~\cite{Rao1988}, and infinite layers~\cite{Siegrist1988} between two neighboring insulating layers. More importantly, typically the bilayered and trilayered materials have a clear higher 
critical transition temperature $T_c$ than monolayered ones~\cite{Rao1988}.
For the FeSCs, however, only monolayered (e.g. 1111 system and 21311 system)~\cite{LaFeAsO,Zhu2009} and infinite-layered systems (e.g. 11 system and 122 system)~\cite{FeSe,122} 
were discovered for a rather long time. Recently, by the intergrowth 
of 1111- and 122-type FeSCs, a series of bilayered compounds 
AB$_2$Fe$_4$As$_4$C$_2$ (A = K, Rb, Cs; B = Ca, Nd, Sm, Gd, Tb, Dy, Ho; C = F, O) were reported with $T_c$ =28-37 K~\cite{12442-1,12442-2,12442-3,12442-4,12442-5}. 
This 12442 system blaze a new trail to explore materials with higher $T_c$ and possible new physical
manifestations of FeSCs. For example, the breaking of the S$_4$ symmetry in the crystal lattice leads to a more complicated band structure with ten Fermi surfaces~\cite{Wang2016}. 

The high-quality single crystals are essential to investigating the intrinsic properties of this system. Up to now, most of the work on this system were carried out based on the
polycrystalline samples~\cite{Ishida2017,Kirschner2018,Adroja2018,Wang2019}, although the CsCa$_2$Fe$_4$As$_4$F$_2$ single crystals have been grown and the gap structure was 
studied by heat transport and lower critical field measurements~\cite{WangCrystal,Huang2019}. The KCa$_2$Fe$_4$As$_4$F$_2$ compound
with a stronger interlayer coupling within a FeAs-K/Cs-FeAs block possesses a higher $T_c$~\cite{12442-1} and more luxuriant properties can be expected. In this paper, with the successful growth of high-quality single crystals of
KCa$_2$Fe$_4$As$_4$F$_2$, we really observed the unusual steep increase of the upper critical field with cooling, suggesting a very robust superconductivity against the external magnetic field near $T_c$. 
Understanding the mechanism of such an extraordinary behavior is a severe challenge and will promote the progress of this field. Moreover, the present finding may have potential values in the applications under 
high fields.

\section{II. Experimental}
Single crystals of KCa$_2$Fe$_4$As$_4$F$_2$ were grown using KAs as the self flux. The raw materials are K chunk (purity 99\%), Ca granules (purity 99.5\%), Fe powders (purity 99+\%), As grains (purity 99.9999\%) 
and CaF$_2$ powders (purity 99.95\%). KAs was synthesized with stoichiometric ratio of K and As in the alumina crucible, which was loaded into a stainless steel pipe container~\cite{Kihou2010} and heated at 650 $^o$C for 10 hours. 
Precursors CaAs and Fe$_2$As were synthesized via solid-state reactions in evacuated quartz tubes by heating the mixed reagents at 700 $^o$C and 750 $^o$C for 12 hours, respectively. 
These precursors and  CaF$_2$ were mixed in an appropriate ratio with excess 15 times KAs as the flux. We found that the crystals of 122 system KFe$_2$As$_2$ was very easy to be produced if the stoichiometric ratio 
was used. Thus we added additional amounts of 100\% CaAs and 50\% CaF$_2$ to restrain the formation of KFe$_2$As$_2$.
The mixtures were put into an alumina crucible, subsequently sealed in a stainless steel pipe~\cite{Kihou2010}. 
The whole preparation process was carried out in a glove box filled with argon. Finally, the stainless steel pipe was placed into a preheated furnace to further suppress the formation of KFe$_2$As$_2$. 
The materials were heated at 980 $^o$C for 20 hours, followed by a slow cooling procedure to 900 $^o$C with a rate of 1.6 $^o$C/h. 
Then the pipe was removed from furnace to cool down rapidly to room temperature. The KCa$_2$Fe$_4$As$_4$F$_2$ single crystals could be separated by washing away KAs flux with deionized water.

The microstructure was examined by scanning electron microscopy
(SEM, Zeiss Supra55). The composition of the single crystals was
checked and determined by energy dispersive x-ray spectroscopy (EDS)
measurements on an Bruker device with the model Quantax200. The
crystal structure and lattice constants of the materials were
examined by a DX-2700 type powder x-ray diffractometer using Cu
K$_\alpha$ radiation. The electrical resistivity was measured on the physical property measurement system
(Quantum Design, PPMS). The magnetic susceptibility measurement was
carried out on the magnetic property measurement system (Quantum
Design, MPMS 3) with the magnetic field oriented parallel to the
ab-plane of the samples.

\section{III. Results and discussion}

\begin{figure}
\includegraphics[width=8.5cm]{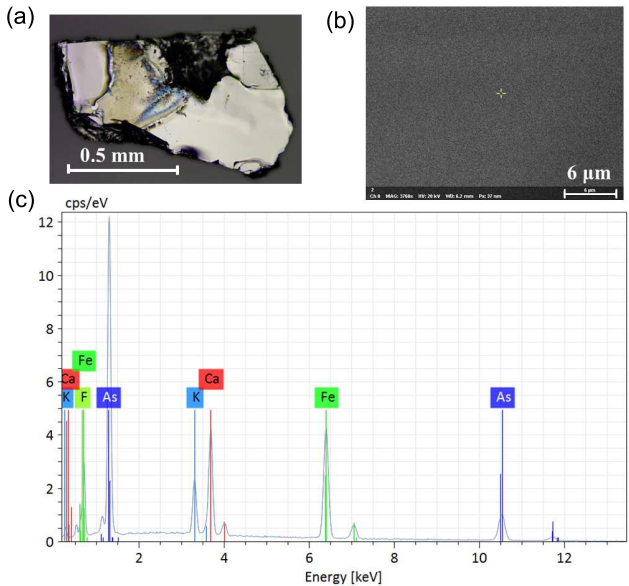}

\caption {(color online)
(a)-(b) The surface pictures of the single crystal taken
with the optical microscope and SEM, respectively. (c) The EDS microanalysis spectrum taken on the surface of the sample. } \label{fig1}
\end{figure}

The morphology of the single crystals was examined by the optical
microscope and the scanning electron microscopy, which are shown in
Figs. 1(a) and (b) respectively. The surface observed from the optical
microscope is shining. The SEM picture shows the clean and flat surface. The typical crystal size was
found to be as large as 1 mm $\times$ 0.8 mm $\times$ 0.06 mm. The composition of the crystals was
examined by the EDS analysis and the typical spectrum is shown in the Fig. 1(c). 
The result of the composition analysis is shown in table
\ref{tab.1}. The ratio of K : Ca : Fe : As : F is 0.94 : 2.18 : 4 : 3.89 : 1.85, which is close to the the expected  1 : 2 : 4 : 4 : 2.

\begin{table}
\centering \caption{Compositions of the KCa$_2$Fe$_4$As$_4$F$_2$ single crystal characterized by EDS measurements.}
\begin{tabular}
{p{1.5cm}<{\centering}p{1.5cm}<{\centering}p{1.8cm}<{\centering}p{1.8cm}<{\centering}}\hline \hline
Element &       Weight   &  Atomic  &  Error(3$\sigma$)  \\
 &        (wt.\%)   &  (at.\%)  &   (\%) \\
\hline
K          & 5.48   & 7.34    &  0.56    \\
Ca          & 12.97   & 16.96    &  1.17    \\
Fe          & 33.13   & 31.09    &  2.61 \\
As          & 43.20    & 30.22     &  3.71 \\
F          & 5.22   & 14.40    &  2.16  \\
 \hline \hline
\end{tabular}
\label{tab.1}
\end{table}

\begin{figure}
\includegraphics[width=9cm]{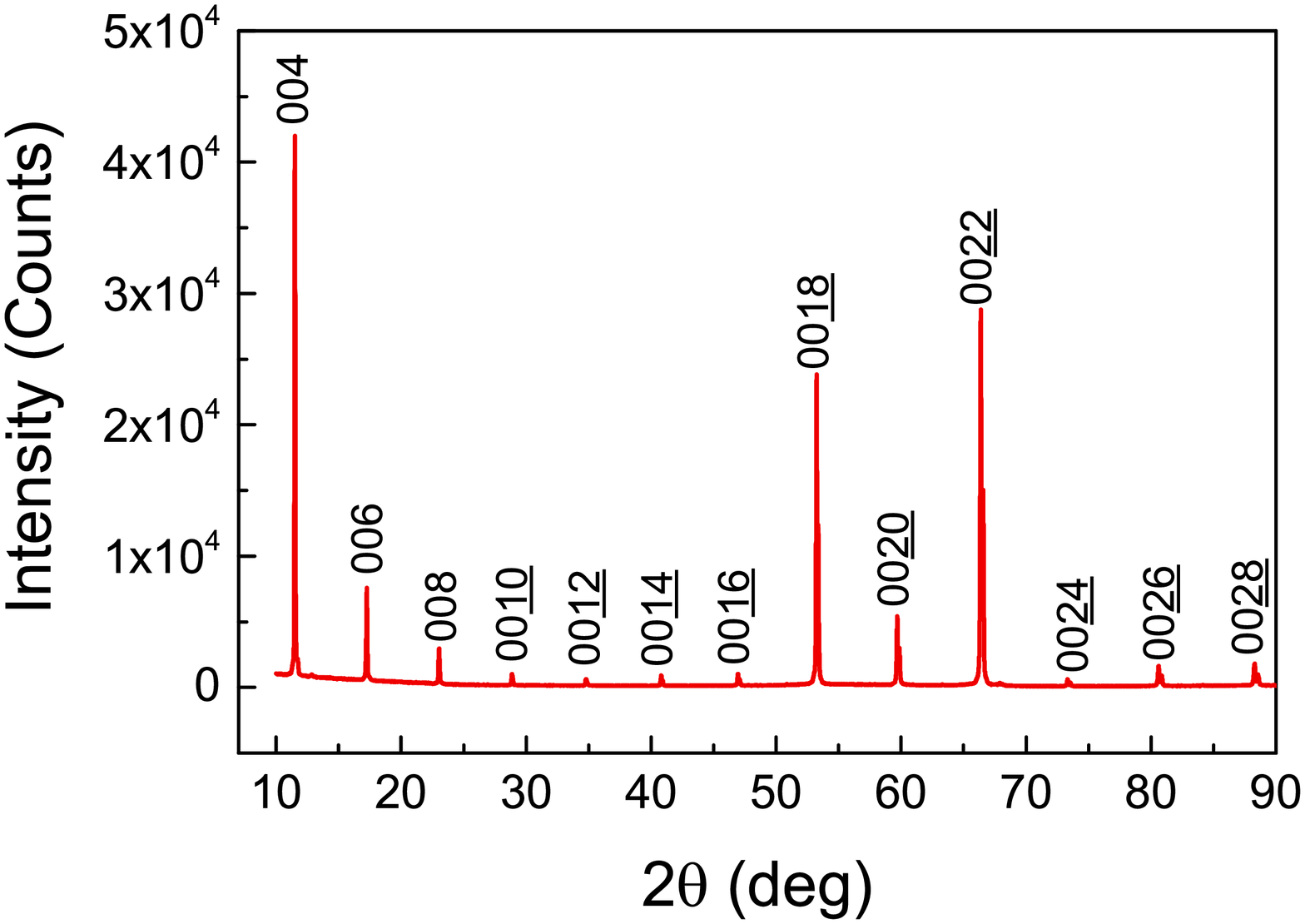}
\caption {XRD patterns of KCa$_2$Fe$_4$As$_4$F$_2$ single
crystal. } \label{fig2}
\end{figure}

The structure of the crystals was 
checked by the x-ray diffraction (XRD) measurement, where the x-ray was incident
on the ab-plane of the crystal. The diffraction patterns are shown in
Fig. 2. All the diffraction peaks can be indexed to the 12442 compound with a tetragonal structure. Only sharp peaks
along (00 2$l$) orientation can be observed, suggesting a high c-axis
orientation. The full width at half maximum (FWHM) of the
diffraction peaks is only about 0.10$^\circ$ after deducting the
$K_{\alpha2}$ contribution, indicating a rather fine crystalline
quality. The c-axis lattice constant was obtained to be 30.991 {\AA}
by analyzing the diffraction data, which is consistent with the previous report on the polycrystalline samples~\cite{12442-1}.

\begin{figure}
\includegraphics[width=8.5cm]{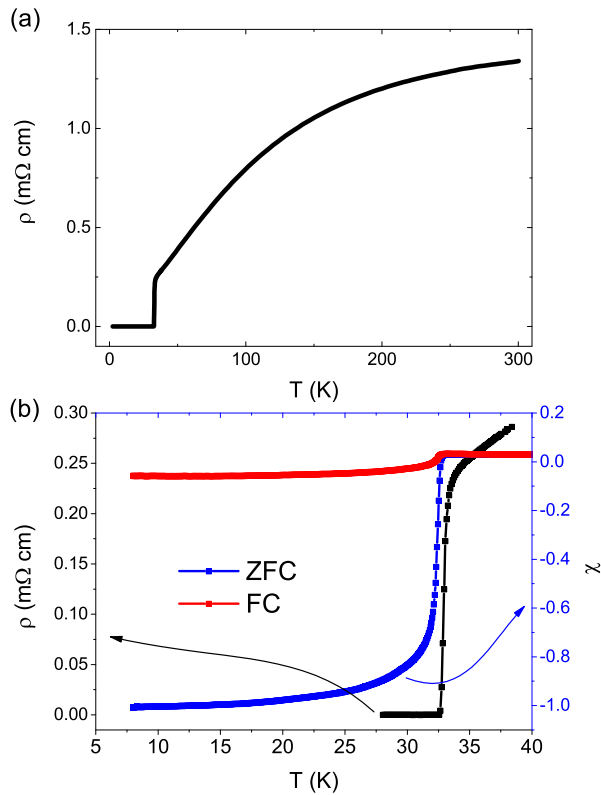}
\caption {(a) Temperature dependence of resistivity measured in
a wide temperature range 0 - 300 K under zero magnetic field. (b)
The magnetic susceptibility measured in zero-field-cooled
(ZFC) and field cooled (FC) models and the resistivity data in the
low temperature range near the superconducting transition.} \label{fig3}
\end{figure}

Temperature dependence of resistivity in the temperature range from
0 to 300 K for is shown in Fig. 3(a). The $\rho-T$ curve shows a clear negative curvature in a wide temperature region before entering the superconducting states,
which is similar with that observed in CsCa$_2$Fe$_4$As$_4$F$_2$~\cite{WangCrystal} and seems to be a common feature for the hole-doped FeSCs~\cite{Mu-1,Mu-2}. 
The onset of the superconducting transition appears at about 33.5 K, whereas the zero resistivity is
reached at about 32.5 K. The dc magnetic susceptibility for the same
sample was measured under a magnetic field of 1 Oe in
zero-field-cooling and field-cooling processes, which is presented
in Fig. 3(b) with temperature between 0 and 40 K. In
order to minimize the effect of the demagnetization, the magnetic
field was applied parallel to the ab-plane of the crystal. The
absolute value of magnetic susceptibility $\chi$ is about 101\%, indicated a high superconducting volume fraction of our samples.
An enlarged $\rho-T$ curve is also shown in figure Fig. 3(b) in
order to have a comparison with susceptibility curve expediently. Both the $\rho-T$ and $\chi-T$ curves display a rather sharp superconducting transition, indicating the high quality of our 
samples. The onset transition temperature revealed by the $\chi-T$ curve is roughly corresponding to the zero resistivity temperature, which is
rather reasonable and common for the compound superconductors.

In order to study the upper critical field $H_{c2}$ and irreversible field $H_{irr}$, we perform the measurements of temperature dependent electronic
resistivity with the magnetic field along two different orientations. As shown in Figs. 4(a) and (b), the
SC transition point shifts to lower temperature with the increase of
the magnetic field for both the orientations: $H \parallel c$ and $H \parallel ab$. In all the measurements, the current was applied with the $ab$ plane and always 
perpendicular to the magnetic field. It is worthy to note that the SC transition for the orientation of $H\parallel c$ shifts
much quicker than that of $H\parallel ab$ by comparing the two sets of data. Remarkably, the superconductivity is very robust against the in-plane field: a magnetic field as high as 9 T
reduces the superconducting transition merely about 0.2 K and 1 K if we check the onset point and zero-resistivity point respectively. 
Quantitatively, we use the criteria $90\%\rho_n$ and $10\%\rho_n$ to
determine the values of $H_{c2}$ and $H_{irr}$ respectively. The temperature
dependence of $H_{c2}$ and $H_{irr}$ is shown in the inset of Fig. 4(b)
for both the two orientations. The vortex-liquid regions between $H_{c2}$ and $H_{irr}$ for the orientation $H\parallel c$ is much larger than that with $H\parallel ab$, in accordance with the 
two-dimensional feature of this material where the pancake-like and Josephson-like vortices will form in the two cases respectively~\cite{Blatter1994}.

\begin{figure}
\includegraphics[width=8.5cm]{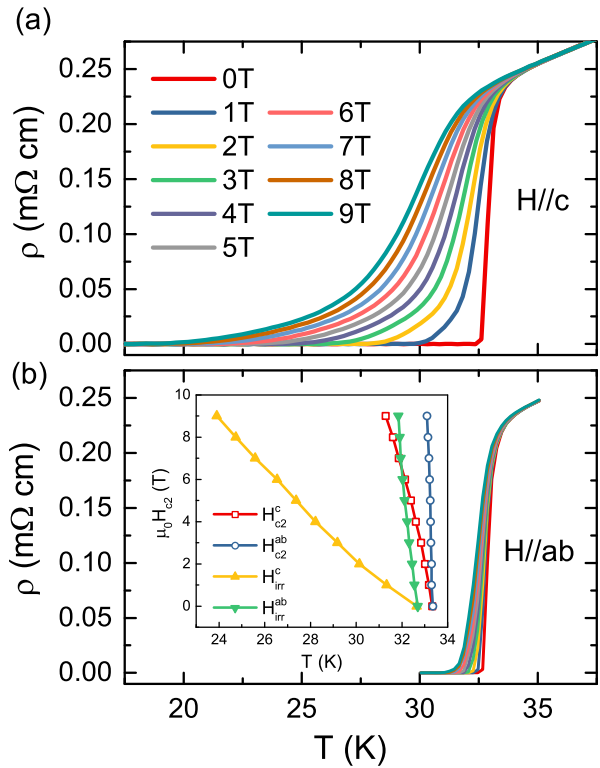}
\caption {(a)-(b) The electronic resistivity as a function of
temperature under the magnetic field up to 9 T with $H//c$ and
$H//ab$, respectively. The inset of (b) shows the upper critical
fields $H_{c2}$ and irreversible field $H_{irr}$ as a function of temperature for two different
orientations. }
\label{fig4}
\end{figure}

The most important issue is about the upper critical field $H_{c2}$. We deduced the slope of the tangent of the $H_{c2}-T$ curves (as shown in the inset of Fig. 4(b)) near $T_c$, 
$d\mu_0H_{c2}/dT$$\mid$$_{T_c}$. We obtained an unexpectedly high value for the $H \parallel ab$ case, $d\mu_0H_{c2}^{ab}/dT$$\mid$$_{T_c}$ = -50.9 T/K. For $H \parallel c$, the slope is -6.4 T/K
which is also large, although not so conspicuous as the in-plane case. These two values give an estimation for the anisotropy of about 8 near $T_c$. 
To have a concise impression, we have summarized the results of different FeSC systems in Table \ref{tab.2}. 
Similar criteria (90\% or 95\%$\rho_n$) were adopted in these reports~\cite{WangCrystal,Jia2008,ZSWang2008,Meier2016}, which facilitates the comparison with our results. One can see that, for the in-plane $H_{c2}^{ab}$, a slope around -10 T/K is a typical value
for various systems of FeSCs including the 1111 system, 122 system, 1144 system, etc. In the 12442 system CsCa$_2$Fe$_4$As$_4$F$_2$, an abnormally steep slope of about -18.2 T/K has been reported previously,  
which revealed preliminarily the peculiarity of this bilayered FeSC system. Now with our result, a slope of -50.9 T/K in the KCa$_2$Fe$_4$As$_4$F$_2$ system further highlights this tendency.
Applying the Werthamer-Helfand-Hohenberg (WHH) relation\cite{Werthamer1966} $\mu_0H_{c2}(0)=-0.693 d \mu_0H_{c2}(T)/dT|_{T_c}
T_c$, we get a vary high value, about 1180 T, for the in-plane upper critical field at zero temperature $\mu_0H_{c2}^{ab}(0)$. Obviously this value is very impressive, although the WHH relation
may overestimated the $\mu_0H_{c2}^{ab}(0)$ in such a multi-band material.  

\renewcommand\arraystretch{1.5}
\begin{table}
\centering \caption{Slope of the upper critical fields with temperature for different systems of FeSCs.}
\begin{tabular}
{p{2.5cm}<{\centering}p{1.5cm}<{\centering}p{1.8cm}<{\centering}p{1.8cm}<{\centering}}\hline \hline
Materials &    $\frac{d\mu_0H_{c2}^{ab}}{dT}$$\mid$$_{T_c}$   &  $\frac{d\mu_0H_{c2}^{c}}{dT}$$\mid$$_{T_c}$  &  Ref.  \\
 &                 (T/K)                                        &  (T/K)  &     \\
\hline
NdFeAsO$_{0.82}$F$_{0.18}$     & -9   & -2.09    &  \cite{Jia2008}    \\
Ba$_{0.6}$K$_{0.4}$Fe$_2$As$_2$    & -9.35   &-5.49    &  \cite{ZSWang2008}    \\
CaKFe$_4$As$_4$          & -10.9   & -4.4    &  \cite{Meier2016} \\
CsCa$_2$Fe$_4$As$_4$F$_2$         & -18.2    & -2.9     &  ~\cite{WangCrystal} \\
KCa$_2$Fe$_4$As$_4$F$_2$          & -50.9   & -6.4    &  This work  \\
 \hline \hline
\end{tabular}
\label{tab.2}
\end{table}

It is difficult to see through the hidden physical mechanism for such an observation at the present stage. Nevertheless, comparing the two 12442 cousins, CsCa$_2$Fe$_4$As$_4$F$_2$
and KCa$_2$Fe$_4$As$_4$F$_2$, can still supply us some clues. The value of $d\mu_0H_{c2}^{ab}/dT$$\mid$$_{T_c}$ for KCa$_2$Fe$_4$As$_4$F$_2$ is about three times as much as that for CsCa$_2$Fe$_4$As$_4$F$_2$. 
Due to the smaller ionic radius of K$^+$ compared with that of Cs$^+$, the distance between the two adjacent
FeAs layers separated by the alkali metals should be shorter for KCa$_2$Fe$_4$As$_4$F$_2$, leading to a stronger interlayer coupling. It is exactly this factor that give rise to
the enhancement of the superconducting transition temperature $T_c$. Thus, it is rather natural and reasonable to speculate that the interlayer coupling within a bilayer 
FeAs-K/Cs-FeAs block is very crucial for the abnormally steep slope of $d\mu_0H_{c2}^{ab}/dT$$\mid$$_{T_c}$. Of course, this needs the verifications of further investigations in the future.
Another clue stems from the comparison with other FeSC systems. As can be seen in Table II, the out-of-plane slope $d\mu_0H_{c2}^{c}/dT$$\mid$$_{T_c}$ of 122 system Ba$_{0.6}$K$_{0.4}$Fe$_2$As$_2$\cite{ZSWang2008} 
is also rather high and only slightly lower than that of KCa$_2$Fe$_4$As$_4$F$_2$. It seems that it is the small anisotropy that restrains the further great enhancement of the in-plane slope in 122 system. 
According to this idea, the present 12442 system reserves the high out-of-plane value of the hole-doped 122 system, and meanwhile achieves a high anisotropy due to the intergrowth with the 1111 component. 
This may explain phenomenologically the possible origin of our observations.
Besides, but not less important, the present system also have a good potential for the applications in high field because of the rather high in-plane irreversible field $H_{irr}^{ab}$.

\section{IV. Conclusions}

In this work, millimeter sized single crystals of
KCa$_2$Fe$_4$As$_4$F$_2$, a bilayered FeSC resembling the famous Bi-2212 system, were successfully grown by a self-flux method.
The chemical compositions, crystal structure, resistivity, and
magnetization susceptibility were investigated systematically. The $T_c$ of the single crystals was confirmed to be about 33.5 K by both
the resistivity and magnetic susceptibility.
The most important, it is found that the slope of the $H_{c2}^{ab}-T$ curve adjoining $T_c$ is very large, indicating an extremely high upper critical field in this system.
Our findings demonstrate that the KCa$_2$Fe$_4$As$_4$F$_2$ material has significant values in the superconducting fundamental research and applications.

\section*{Acknowledgments}
This work is supported by the Natural Science Foundation of China
(No. 11204338), the ``Strategic Priority Research Program (B)" of
the Chinese Academy of Sciences (No. XDB04040300 and XDB04030000)
and the Youth Innovation Promotion Association of the Chinese
Academy of Sciences (No. 2015187).

%

\end{document}